\documentclass[aps,pra,preprint,groupedaddress,floatfix]{revtex4-1}

\usepackage{graphicx}
\usepackage{hyperref}
\usepackage{amsmath,amsthm,amssymb}
\usepackage{booktabs}

\begin{document}

\title{Calculation of Stark resonance parameters for valence orbitals
  of the water molecule.}


\author{Susana Arias Laso}
\author{Marko Horbatsch}
\affiliation{Department of Physics and Astronomy, York University,
  Toronto, Ontario, Canada M3J 1P3}


\date{\today}

\begin{abstract}
  An exterior complex scaling technique is applied to compute Stark
  resonance parameters for two molecular orbitals ($1b_{1}$ and
  $1b_{2}$) represented in the field-free limit in a single-center
  expansion. For electric DC field configurations that guarantee
  azimuthal symmetry of the solution the calculation is carried out by
  solving a two-dimensional partial differential equation in spherical
  polar coordinates using a finite-element method. The resonance
  positions and widths as a function of electric field strengths are
  shown for field strengths starting in the tunnelling ionization
  regime, and extending well into the over-barrier ionization region.
\end{abstract}


\maketitle

\section{\label{sec:intro} Introduction}

The study of water molecule vapour in strong fields has been an area
of recent interest, particularly in the field of intermediate-energy
ion-molecule collisions~\cite{PRA.85.052713, PRA.86.022719,
  PRA.93.052705, PRA.93.032704, PRA.93.062706, PRA.87.032709,
  PRA.87.052710}, where capture and direct ionization processes
compete. Most of the theoretical studies are in the context of an
independent-electron approximation which involves a molecular orbital
(MO) representation. First studies of the properties of molecular
orbitals exposed to laser pulses have also been
performed~\cite{PRL.107.083001, PRA.81.023412}.

For the problem of strong electric DC fields numerous investigations
have been carried out for the hydrogen molecular ion, where
interesting structures occur in the resonance width as a function of
internuclear separation~\cite{JPB.46.085004, PRA.63.013414,
  JPB.29.4625}. This also carries over to the case of low-frequency AC
fields, i.e., infrared laser fields with the help of the Floquet
method~\cite{JPB.46.245005}. Experimentally, the investigation of
water vapour is challenging, but possible~\cite{PRL.74.1962,
  PRL.86.3751, PRA.70.062716, PRA.75.042711}. Due to the importance of
the water molecule in applied fields (e.g., radiation therapy) one
should expect more work to be carried out in this research area in the
near future.

From the point of view of a theoretical description, the subject is
challenging due to the multi-center nature of the combined Coulomb
interactions. Thus, many investigations in the context of
electron/positron or ion scattering are carried out in single-center
approximations for the molecular target. For a number of situations
this approach appears to be successful in that ionization (and
capture) cross sections are obtained which agree reasonably with
experiment. This is in part the case since experiments are usually not
sensitive to the orientation of the molecule during the collision.

A satisfactory description of the molecular structure of H$_2$O was
obtained within the independent-electron approximation by the
self-consistent field (SCF), or variational Hartree-Fock method with
multi-center Slater orbitals~\cite{PitzerJCP49}. The direct
application of these orbitals for collisional or strong-field studies
does represent significant computational and methodological
challenges. An application of the SCF method using a single-center
Slater basis has been available~\cite{MocciaJCP40I, MocciaJCP40II,
  MocciaJCP40III}, and has been used in some collision
studies~\cite{JPB.47.015201}. The molecular orbitals for water
from~\cite{MocciaJCP40III} have been compared to experimental electron
spectroscopy studies~\cite{CPL.439.55}, and also to more sophisticated
calculations, and were found to give reasonable agreement with
observations. For further studies of state-of-the art spectrocopy and
calculations we refer the reader to~\cite{Ning200819}.

In order to use a variational SCF solution for collisional or
strong-field perturbation studies one faces the need to define a
consistent molecular Hamiltonian. In the present work we limit
ourselves to two of the three molecular valence orbitals of the water
molecule where the wave functions are dominated by a single angular
momentum symmetry ($1b_{1}\approx 2p_x$ or $1b_{2}\approx 2p_y$); here
we assume that the protons are in the $y-z$ plane, as shown in
Fig.~\ref{fig:orbitals}. For our approximate treatment it is
straightforward to obtain an effective single-electron potential for
each orbital. The calculation of Stark resonance parameters for these
approximate single-center molecular orbitals does represent a first
step to be followed by more ambitious modelling to be carried out in
the future.

We present a complex scaling approach to study the effect of an
external DC field on the $1b_{1}\approx 2p_{x}$ and $1b_{2}\approx
2p_{y}$ MO's of H$_{2}$O, where the problem is expressed as a system
of coupled partial differential equations (PDEs) with an effective
potential that reflects the binding properties of each orbital. The
paper is organized as follows: The problem as formulated in spherical
polar coordinates is introduced in Sec.~\ref{sec:pde_approach}, with
some technical details concerning the construction of the electronic
potential being described in Sec.~\ref{sec:Veff}. The exterior complex
scaling formalism and the implementation are explained in
Sec.~\ref{sec:ecs}. The numerical results are presented in
Sec.~\ref{sec:results}, followed by conclusions in
Sec.~\ref{sec:conc}. Atomic units ($\hbar = m_{e} = e =
4\pi\epsilon_{0} = 1$) are used throughout.

\section{\label{sec:pde_approach} PDE approach to the problem in spherical
  polar coordinates}

The set of single-center wave functions introduced by
Moccia~\cite{MocciaJCP40I, MocciaJCP40II, MocciaJCP40III} is taken as
a starting point in the present approach. The general expression for
the basis functions is defined by a Slater-type
orbital~\cite{MocciaJCP40I}
\begin{widetext}
\begin{eqnarray}
 f_{n, l, m}(\zeta,r,\theta,\phi) & = & \sqrt{\frac{(2\zeta)^{2n+1}}{(2n)!}}
r^{n-1}\exp(-\zeta r) S_{l, m}(\theta,\phi),
\label{sto}
\end{eqnarray}
\end{widetext}
where the angular part $S_{l,m}(\theta,\phi)$ represents real
spherical harmonics. The expansion coefficients and non-linear
coefficients $\{\zeta_{i}\}$, determined by Roothaan's
self-consistent-field procedure~\cite{RoothaanSCF, MocciaJCP40III},
were used to construct a reduced form of the radial functions that
describe all the molecular orbitals, and in particular the $1b_{1}$
and $1b_{2}$ states. In Fig.~\ref{fig:Veff1b11b2} we depict
schematically how the orbitals are approximated as $|m| = 1$
eigenstates of spherically symmetric potentials. For the purpose of
this study, the expansion of the orbital functions was truncated at
$n=2$ with their associated $(n, l, m)$ combinations.

The framework for the present study involves the construction of the
electronic potential as an effective orbital-dependent potential
$V_{\rm{eff}}(r)$ extracted from the single-center Moccia wave
functions. Then we apply exterior complex scaling (ECS) to determine
the numerical solution of the problem associated with a molecular
orbital in the presence of a strong electric DC field applied along
the $z-$direction. A partial differential equation (PDE) needs to be
solved to determine the resonance position and width.

The Schr\"{o}dinger equation for the bound-state problem is expressed
in spherical polar coordinates as
\begin{eqnarray}
  [-\frac{1}{2}\frac{d^2}{dr^2}+\frac{\hat{L}^2}{2r^2}+V_{\rm{eff}}(r)]\psi
  & = & E\psi,
\label{sch_eq_noCS}
\end{eqnarray}
where $\hat{L}^{2}$ is the orbital angular momentum operator.

\begin{figure}[t!]
\includegraphics[width=0.25\textwidth]{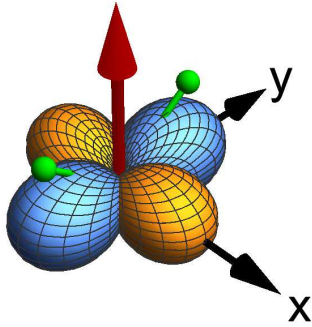}
\caption{Schematic display of the $1b_1\approx 2p_x$ (shown in yellow)
  and $1b_2\approx 2p_y$ (shown in blue) molecular orbitals. Also
  indicated (in green) is the orientation of the protons. The $z$-axis
  (in red) is the direction of the external electric field of strength
  $F_0$.}
\label{fig:orbitals}
\end{figure}

Figure~\ref{fig:orbitals} shows a scheme of the geometry of the
system, where the orientation of the $1b_{1}$ and $1b_{2}$ MO's is
represented with respect to the plane where the protons are located.
The direction of the applied electric field along $\hat{z}$ is
included as well.

\subsection{\label{sec:Veff} Construction of an electronic effective potential}

The effective potential is obtained in two steps: first the orbital
energy and the STO are inserted into the single-electron
Schr\"{o}dinger equation~(\ref{sch_eq_noCS}) which is then solved for
$V_{\rm{eff}}^{(1)}(r)$. In the second step we apply the so-called
Latter correction~\cite{PhysRev.99.510} to ensure that the asymptotic
behavior of the potential is correct, i.e., proportional to $-1/r$.
\begin{eqnarray}
V_{\rm{eff}}(r) = \left.
\begin{cases}
V_{\rm{eff}}^{(1)}(r) & \text{for } r < r_{0} \\
-1/r & \text{for } r > r_{0}
\end{cases}
\right\}
\label{coulomb_tail}
\end{eqnarray}
The point $r_{0}$ is determined from
$V_{\rm{eff}}^{(1)}(r_{0}) = -1/r_{0}$ and is found to be sufficiently
large that the original self-consistent field orbital energy used to
derive $V_{\rm{eff}}^{(1)}(r)$ is close to the eigenenergy
of~(\ref{sch_eq_noCS}), with at least two significant digits of
agreement, with $V_{\rm{eff}}(r)$ given by~(\ref{coulomb_tail}).

\begin{figure}[t!]
\includegraphics[width=0.49\textwidth]{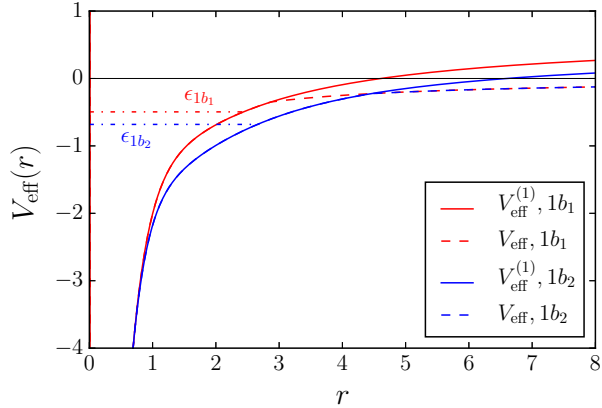}
\caption{Electronic effective potential in atomic units for the
  $1b_{1}\approx 2p_{x}$ (red) and $1b_{2}\approx 2p_{y}$ (blue) MO's
  of the H$_{2}$O molecule. The solid lines give the potential as
  derived from~(\ref{sch_eq_noCS}) using the SCF orbitals and
  eigenenergies, while the dashed lines show the potentials after the
  Latter correction is applied. The dot-dashed lines show the
  eigenenergies obtained from the Moccia wave
  functions~\cite{MocciaJCP40III}.}
\label{fig:Veff1b11b2}
\end{figure}

In Figure~\ref{fig:Veff1b11b2}, we show a comparison of the effective
potential $V_{\rm{eff}}^{(1)}(r)$ (continuous line) derived from the
Moccia wave functions representing the $1b_{1}$ and $1b_{2}$
MO's~\cite{MocciaJCP40III}, and the transformed electronic potential
$V_{\rm{eff}}(r)$ (dashed line) after the Latter correction was
implemented. The effective potentials for the $1b_{1}$ and $1b_{2}$
MO's are given as the more shallow and deeper curves. As
Figure~\ref{fig:Veff1b11b2} illustrates, one drawback of the method is
that the effective potential is orbital dependent. A direct
consequence is that the value of $r=r_{0}$, which sets the position in
$r$ where the Coulombic tail is imposed, differs between the MO's,
being almost twice as large for the $1b_{2}$ as compared to the
$1b_{1}$ MO.

\subsection{\label{externalDC} H$_{2}$O in an external electric DC field}

Now that we have obtained an effective potential to define the
field-free Schr\"{o}dinger equation~(\ref{sch_eq_noCS}) for an orbital
obtained in the SCF method we proceed with the problem of the molecule
exposed to a strong DC field.

When an electric field is applied in the $\hat{z}$ direction,
$\vec{F} = F_{0}\hat{z}$, the separation-of-variables ansatz as
applied to the Schr\"{o}dinger equation
\begin{eqnarray}
\psi(r,\theta,\phi) & = & \psi(r,\theta)\exp(im\phi)
\label{sov}
\end{eqnarray}
leads to a PDE that represents the Stark problem for an H$_{2}$O
orbital:
\begin{widetext}
\begin{eqnarray}
  -\frac{1}{2}\frac{\partial^{2}\psi}{\partial r^2}-\frac{1}{2r^2}
  (\frac{\cos\theta}{\sin\theta}\frac{\partial\psi}{\partial\theta}+
  \frac{\partial^2\psi}{\partial\theta^2})
  +(\frac{m^2}{2r^2\sin^2\theta}+V_{\rm{eff}}(r)-E +
  F_{0}r\cos\theta)\psi & = & 0.
\label{pde}
\end{eqnarray}
\end{widetext}

Here the complex eigenenergy $E$ contains the information about the
resonance position (real part), i.e., $E_{R}$ and width $\Gamma$
(imaginary part is $-\Gamma/2$), and may be expressed as
\begin{eqnarray}
E & = & E_{R} + iE_{I} = E_{R} - i\frac{\Gamma}{2},
\label{complex_E}
\end{eqnarray}
The imaginary part $\Gamma$ is related to the lifetime of the decaying
state $\tau$ via $\Gamma\tau = 1$. For the $1b_{1}$ and $1b_{2}$
orbitals we have $m=\pm 1$ respectively. The presence of the effective
potential $V_{\rm{eff}}(r)$ makes this problem challenging in the
sense that it is not possible to obtain separable solutions, as for
the hydrogen atom in which a pure Coulomb potential leads to
separability in parabolic coordinates~\cite{TelnovJPB22}. It is then
necessary to generate a more general solution by solving the PDE
numerically, e.g., by applying a finite-element method.

\subsection{\label{sec:ecs} Exterior complex scaling}

The ionization regime of the water molecule will be described by means
of a non-hermitian Hamiltonian that reveals discrete resonance
eigenvalues containing information about the quasi-bound states that
tunnel through the barrier or escape over the potential barrier for
strong fields. Among the different techniques implemented to compute
the resonance energies established methods are the complex
scaling~\cite{complexScaling, complexScalingBaslev,
  complexScalingSimon}, and exterior complex scaling~\cite{ecsSimon};
the latter was introduced as an extension of the former method. These
have been widely used in scattering
problems~\cite{complexScalingBaslev, complexScalingSimon}, and also in
time-dependent Schr\"{o}dinger equation problems for strong
fields~\cite{ecsRuiz, ecsTao, ecsScrinzi}. For our aim of studying the
field ionization properties of H$_{2}$O orbitals, we implement a
modified exterior complex scaling technique in which the radial
coordinates are extended into the complex plane by a phase factor,
which is turned on gradually beyond some distance from the
origin. This method allows to address the tunneling and over-barrier
ionization problem by avoiding the complication of describing
quasi-bound states with outgoing waves for $r\to\infty$.

In the present work the complex scaling transformation is given by
\begin{eqnarray}
r & \rightarrow & r\exp(i\chi(r)),
\label{ecs_r}
\end{eqnarray}
where $\chi(r)$ is defined as a function of the $r$ coordinate with
the purpose of making the scaling gradually effective from some
vicinity of $r=r_{\rm{s}}$ on,
\begin{eqnarray}
\chi(r) & = & \frac{\chi_{\rm{s}}}{1+\exp[-\frac{1}{\Delta r}(r-r_{\rm{s}})]}.
\label{ecs_theta}
\end{eqnarray}
For given $r_{s}$ one has to choose $\Delta r$ sufficiently small, so
that the function $\chi(r)$ starts from small values at $r = 0$. For
large $r$ it reaches the value $\chi_{\rm{s}}$.

The set of possible values for the asymptotic scaling angle
$\chi_{\rm{s}}$ and the parameters $r_{s}$ and $\Delta r$, which
control where and how quickly the scaling is turned on, was explored
in detail in order to establish how sensitive the PDE solutions are,
and to test the effectiveness of the complex scaling technique to
absorb the outgoing wave. Numerous tests were carried out to ensure
that the ``exact'' results of Telnov~\cite{TelnovJPB22} for atomic
hydrogen orbitals including $2p$ were reproduced.

In order to investigate the effects of the DC field on the H$_{2}$O
orbital energies it is necessary to consider the extra terms that the
exterior complex scaling~(\ref{ecs_r}) introduces in the
Schr\"{o}dinger equation~(\ref{pde}). Aditionally, we need to turn the
scaling on only in the regime $r>r_{0}$ (Eq.~(\ref{coulomb_tail}))
such that we have a simple Coulomb potential in the scaling region. In
order to make use of standard finite element methods, the
complex-valued wave function is separated into real and imaginary
parts, such that a system of coupled differential equations is
obtained as follows:
\begin{widetext}
\begin{eqnarray}
  -\frac{1}{2}\frac{\partial^{2}\psi_{R}}{\partial r^2}-\frac{1}{2r^2}
  (\frac{\cos\theta}{\sin\theta}\frac{\partial\psi_{R}}{\partial\theta}+
  \frac{\partial^{2}\psi_{R}}{\partial\theta^2}) \nonumber\\
  +(\frac{m^2}{2r^2\sin^2\theta}+V_{\rm{eff}}^{R}(r)c_{2}-V_{\rm{eff}}^{I}(r)s_{2}
  -E_{R}c_{2}+E_{I}s_{2}+
  F_{0}r\cos\theta c_{3})\psi_{R} \nonumber\\
  +(-V_{\rm{eff}}^{R}(r)s_{2}-V_{\rm{eff}}^{I}(r)c_{2}+E_{R}s_{2}+E_{I}c_{2}-F_{0}r\cos\theta
  s_{3})\psi_{I} & = & 0, \nonumber\\
  \vspace{1cm}
  -\frac{1}{2}\frac{\partial^{2}\psi_{I}}{\partial r^2}-\frac{1}{2r^2}
  (\frac{\cos\theta}{\sin\theta}\frac{\partial\psi_{I}}{\partial\theta}+
  \frac{\partial^{2}\psi_{I}}{\partial\theta^2}) \nonumber\\
  +(\frac{m^2}{2r^2\sin^2\theta}+V_{\rm{eff}}^{R}(r)c_{2}-V_{\rm{eff}}^{I}(r)s_{2}
  -E_{R}c_{2}+E_{I}s_{2}+
  F_{0}r\cos\theta c_{3})\psi_{I} \nonumber\\
  +(V_{\rm{eff}}^{R}(r)s_{2}+V_{\rm{eff}}^{I}(r)c_{2}-E_{R}s_{2}-E_{I}c_{2}+F_{0}r\cos\theta
  s_{3})\psi_{R} & = & 0.
\label{pde_system}
\end{eqnarray}
\end{widetext}
The labels $R(I)$ stand for real and imaginary parts respectively,
also the notation $\{c_{k},s_{k}\}$ is introduced to represent
$\{\cos(k\chi(r)),\sin(k\chi(r))\}$ respectively, with $k = 2, 3$ and
$\chi(r)$ given in~(\ref{ecs_theta}).

The PDE system~(\ref{pde_system}) is solved numerically on a
rectangular mesh defined by the $(r,\theta)$ coordinates, which take
values in the domains $[\epsilon, r_{\rm{max}}]$ and
$[\eta, \pi-\eta]$, respectively. The parameters $\epsilon$ and $\eta$
that define the coordinate ranges, were chosen to be of the order of
$10^{-3}\ \rm{a.u.}$, and the $r-$coordinate extends to
$r_{\rm{max}} = 20\ \rm{a.u.}$ In order to find a correct set of
$\psi_{R(I)}(r, \theta)$ solutions it is essential to impose proper
boundary conditions that ensure the wave functions vanish at the
limits of the mesh. For the $|m|=1$ states we impose the condition
$\psi_{R(I)}(\epsilon, \theta) = \epsilon\sin(\theta) = \epsilon
P_{1}^{1}(\theta)$
which is consistent with the assumption that at small $r = \epsilon$
the lowest term in an expansion in associated Legendre polynomials
dominates and behaves like $Ar^{2}\sin(\theta)$.

We implemented a two-parameter root search for $\{E_{R}, E_{I}\}$ by
solving the PDE as if it were an inhomogeneous problem. We pick a
location in the $(r, \theta)$ plane where the probability amplitude is
expected to be large and vary $\{E_{R}, E_{I}\}$, i.e., effectively
the complex energy $E$ to maximize the amplitude.

\section{\label{sec:results} Stark resonance parameters}

We explored the influence of a set of parameters involved in the 2D
problem~(\ref{pde_system}) on the eigenvalue $E_{R}+iE_{I}$ which
describes the ionization process as an exponential decay in time in
terms of resonance position and half-width. In addition to testing the
code against known results for atomic hydrogen~\cite{TelnovJPB22}, we
have performed systematic studies of our results for the H$_{2}$O
orbitals against a number of parameters in order to assess the
accuracy of these results. One parameter concerns the limiting
resolution with which the finite-element method proceeds (the
MaxCellSize parameter in the Mathematica10 implementation of NDSolve,
which we call $\Delta$).  For values $\Delta<0.02\ \rm{a.u.}$ we find
stability in the eigenvalues (real and imaginary parts) of two-three
significant digits. For the results quoted below we then applied the
more stringent criterion of $\Delta=0.01\ \rm{a.u.}$

The second parameter which was investigated is the range where the
complex scaling function sets in, i.e., $r_{s}$ and $\Delta r$ in
Eq.~(\ref{ecs_theta}). For the scaling method to work we require
scaling to set in for $r>r_0$ when the effective potential represents
a simple Coulomb tail, which in practice is satisfied by
$r_{s}>2r_{0}$. We also need $\Delta r < 2\ \rm{a.u.}$ to guarantee
smooth turn-on in this region. Small values of $\Delta r$ pose
challenges for the automated finite-element method, since in the limit
of $\Delta r \to 0$ one would need to implement the derivative
discontinuity in the solution as discussed by
Scrinzi~\cite{ecsScrinzi}. We find stable results for the real and
imaginary parts of the eigenenergies at the level of three significant
digits for the range $10 < r_{s} < 15\ \rm{a.u.}$ Larger values would
require an increase in the computational domain beyond
$r=20\ \rm{a.u.}$

Finally, another systematic that was explored is the choice of the
ultimate scaling angle reached at large $r$, namely the value of
$\chi_{\rm{s}}$ in~(\ref{ecs_theta}). For an accuracy demand of three
significant digits, and the other parameters chosen in the ranges
described above stability in the resonance widths is achieved for
$0.6< \chi_{\rm{s}} < 1.2$ radians.

\begin{figure}[t!]
\includegraphics[width=0.49\textwidth]{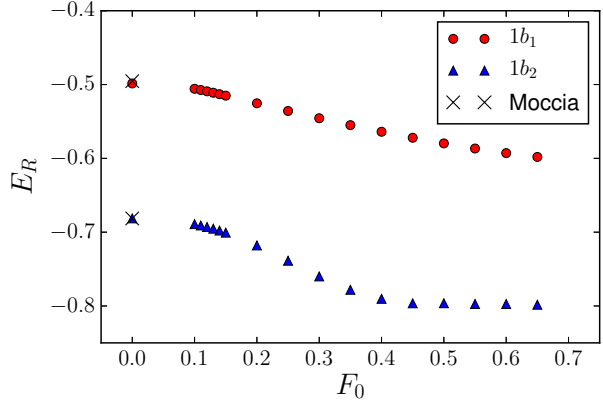}
\caption{Resonance position in atomic units as a function of the
  external field strength $F_{0}$ in atomic units for the $1b_{1}$
  (red circles) and $1b_{2}$ (blue triangles) MO's of H$_{2}$O.}
\label{fig:position1b11b2}
\end{figure}

\begin{figure}[t!]
\includegraphics[width=0.49\textwidth]{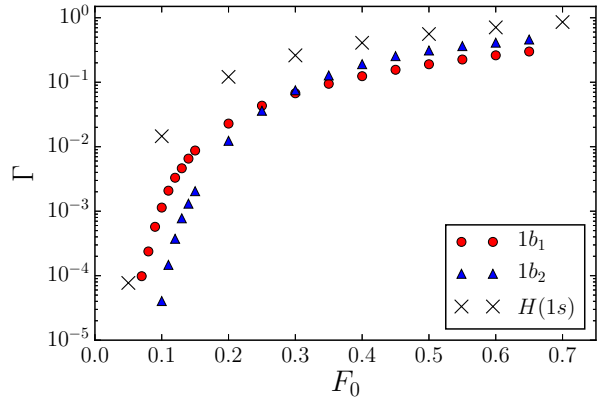}
\caption{Resonance width in atomic units as a function of the external
  field strength $F_{0}$ in atomic units for the $1b_{1}$ (red
  circles) and $1b_{2}$ (blue triangles) MO's of H$_{2}$O. For
  comparison, atomic hydrogen H($1s$) ionization rates from
  Refs.~\cite{TelnovJPB22, KolosovJPB20} are shown as crosses.}
\label{fig:width1b11b2}
\end{figure}

In Fig.~\ref{fig:position1b11b2} we show the resonance position
$E_{R}$ as obtained from the present calculations for the weakly bound
$1b_{1}$ and the strongly bound $1b_{2}$ valence orbitals as a
function of applied electric field strength $F_{0}$. In the limit of
zero field the calculation reproduces the SCF eigenvalues of
Moccia~\cite{MocciaJCP40III}. The field has to be strong (in
comparison with atomic hydrogen results for $2p$
orbitals~\cite{TelnovJPB22}) in order to change the resonance position
appreciably. For the more deeply bound $1b_{2}$ orbital the shift in
resonance position saturates with field strength.

In Fig.~\ref{fig:width1b11b2} the resonance widths are shown for both
orbitals as functions of external field strength $F_{0}$. The graphs
display threshold behavior at the weaker field strengths. As expected
we find a lower threshold (critical field strength) for the more
weakly bound $1b_{1}$ orbital. Interestingly, however at a field
strength of about $F_{0}=0.3\ \rm{a.u.}$ the values for the widths
cross, i.e., the deeper bound $1b_{2}$ orbital displays a larger
ionization rate as the field strength is increased further.

Also shown in Fig.~\ref{fig:width1b11b2} are the widths for the
H($1s$) orbital from Refs.~\cite{TelnovJPB22, KolosovJPB20}. They can
be compared to the $1b_{1}$ orbital results, since the binding energy
is very close in the free-field limit. Since the tunneling barrier is
mostly in the asymptotic regime where the potential energy has a
$-1/r$ tail, it is not surprising that the widths for
H$_{2}$O($1b_{1}$) and H($1s$) share some similarity in shape. In the
tunneling region H($1s$) has an ionization rate that is larger by
about an order of magnitude. In the over-barrier regime, however, the
ionization rates come to within a factor of 3. Reasons for why he
$1b_1$ water molecular orbital is harder to ionize than H($1s$) have
to do with the different shape of the orbital density ($m=1$ \emph{vs}
the spherical H($1s$) density), and the substantially more attractive
potential at shorter distances.

An examination of contour plots of the densities $\Psi^* \Psi$, as
well as of the potential energies $V_{\rm{eff}}(r) - F_{0}z$ for
different field strengths, (both as a function of $r, \theta$), allows
to make the following observations. For field strengths
$F_{0}<0.1\ \rm{a.u.}$ there is a barrier the electrons need to
penetrate in order to be ionized. From the potential energy plot shown
in Fig.~\ref{fig:position1b11b2} one can see that for weak fields
(small values of $F_{0}$) the barrier is longer for the more deeply
bound $1b_{2}$ orbital. This explains why the ionization threshold
occurs for $F_{0}>0.1\ \rm{a.u.}$ for this orbital, which is about a
factor of two larger than for the $1b_{1}$ orbital.

The field strength region where the ionization rates (resonance
widths) display a change in character, i.e., turn over to rise much
more gradually with the field strength $F_{0}$ can be characterized as
a regime where there is a narrow potential saddle at small $\theta$,
in the vicinity of $r\approx 3\ \rm{a.u.}$, such that electron flux
can leave, and is then accelerated by the electric field. The crossing
of the ionization rates for the $1b_{1}$ and $1b_{2}$ orbitals occurs,
since the saddle in the potential becomes effectively lower at strong
fields for the $1b_{2}$ orbital. This can be inferred from the
comparison of the two effective potentials, which share the same
asymptotic behavior beyond $r=4.3\ \rm{a.u.}$ (see
Figure~\ref{fig:Veff1b11b2}).

The origin for the different radial dependencies of the effective
potential for the two orbitals can be found in the geometry of the
water molecule. The weakly bound $1b_{1}$ orbital has its lobes
perpendicular to the plane defined by the location of the three
nuclei. Therefore, it is least affected by the two protons. The
$1b_{2}$ orbital explores the potentials due to the protons more
strongly in the SCF calculation of Moccia, and, therefore, the
resulting (in our approximation spherically symmetric)
$V_{\rm{eff}}(r)$ has a more attractive region in the range
$0.7 < r < 4.3$.

\begin{table}[t]
\centering
\caption{\label{tab:E_values} Resonance positions and widths for different
  field strengths in atomic units. The numbers in brackets indicate the exponent
  $k$, i.e., the numbers are to be multiplied by $10^k$.}
\begin{ruledtabular}
\begin{tabular}{rrrrr}
\toprule
 & & $1b_{1}$ & & $1b_{2}$ \\
$F_{0}$ & $E_{R}$ & $\Gamma$ & $E_{R}$ & $\Gamma$ \\
\midrule
$0.1$ & $-0.506$ & $1.14(-3)$ & $-0.689$ & $4.04(-5)$ \\
$0.2$ & $-0.525$ & $2.28(-2)$ & $-0.718$ & $1.23(-2)$ \\
$0.3$ & $-0.546$ & $6.74(-2)$ & $-0.760$ & $7.51(-2)$ \\
$0.4$ & $-0.564$ & $1.24(-1)$ & $-0.790$ & $1.91(-1)$ \\
$0.5$ & $-0.580$ & $1.90(-1)$ & $-0.796$ & $3.11(-1)$ \\
$0.6$ & $-0.593$ & $2.61(-1)$ & $-0.797$ & $4.11(-1)$ \\
\bottomrule
\end{tabular}
\end{ruledtabular}
\end{table}

We summarize the results in Table~\ref{tab:E_values} for further
reference, i.e., for future comparison with calculations based on
other models for the molecular orbitals.

\section{\label{sec:conc} Conclusion}

We have carried out a study of two of the three valence orbitals of
the water molecule, $1b_{1}$ and $1b_{2}$, in the presence of an
external electric DC field. The tunneling ionization and over-barrier
ionization regimes were explored by finding a numerical solution to
the PDE system defined by an effective potential obtained from
single-center Slater-type orbitals. The exterior complex scaling
parameters as well as a finite-element resolution parameter were
optimized to guarantee a minimum of $2-3$ significant digits for the
solutions. The resonance parameters that describe the ionization
process, resonance position and width, were explored over a wide range
of electric field strengths. We demonstrated how an increase of the
field strength beyond a critical point in the over-barrier region led
to a crossing between the ionization rates of the two orbitals.
Additional observations of the effective potential for different field
strengths were carried out to shed some light on the interpretation of
this behavior.

\bibliography{report}

\end{document}